\documentclass[aip,cha,number,reprint,nofootinbib,a4paper,superscriptaddress,12pt]{revtex4-1}
\usepackage{graphics}
\usepackage{epsfig}
\usepackage{latexsym}
\usepackage{amsfonts}
\usepackage{hyperref}
\usepackage{amsmath}
\usepackage{bm}
\usepackage{dcolumn}
\usepackage{amssymb}
\usepackage{color}
\begin{document}

\title{Components of multifractality in the Central England Temperature anomaly series}

\author{Jeferson \surname{de Souza}}
\email[E-mail address: ]{jdesouza@ufpr.br}
\affiliation{Departamento de Geologia, Universidade Federal do Paran\'{a} \\ PO Box 19027, 81531-980, Curitiba-PR, Brazil}
\affiliation{Centro  Estadual de Educa\c{c}\~{a}o Profissional de Curitiba \\ Rua Frederico Maurer 3015, 81670-020 Curitiba-PR, Brazil}

\author{S\'{i}lvio M. \surname{Duarte Queir\'{o}s}}
\email[E-mail address: ]{sdqueiro@gmail.com}
\affiliation{Centro de F\'{\i}sica do Porto,  Universidade do Porto \\ Rua do Campo Alegre 687, 4169-007 Porto, Portugal}
\affiliation{Istituto dei Sistemi Complessi - CNR \\ Via dei Taurini 19, 00185 Roma, Italy}
\affiliation{Centro Brasileiro de Pesquisas F\'{\i}sicas and \\
National Institute of Science and Technology for Complex Systems\\ Rua Dr Xavier Sigaud 150, 22290-180 Rio de Janeiro - RJ, Brazil}

\author{Alice M.  \surname{Grimm}}
\email[E-mail address: ]{grimm@fisica.ufpr.br}
\affiliation{Departamento de F\'{\i}sica, Universidade Federal do Paran\'{a} \\ PO Box 19044, 81531-990, Curitiba-PR, Brazil}


\begin{abstract}

We study the multifractal nature of the Central England Temperature (CET) anomaly, a time series that spans more than 200 years.
The data are analyzed in two ways: as a single set and by using a sliding window of 11 years. In both cases, we quantify
the width of the multifractal spectrum as well as its components, which are defined by the deviations from the Gaussian distribution
and the dependence between measurements. The results of the first approach show that the key contribution to the multifractal
structure comes from the dynamical dependencies, mainly weak ones, followed by a residual contribution of the deviations from
the Gaussian. The sliding window approach indicates that the peaks in the evolution of the non-Gaussian contribution occur almost
at the same dates associated with climate changes that were determined in previous works using component analysis methods. Moreover,
the strong non-Gaussian contribution from the 1960s onwards is in agreement with global results recently presented.

\end{abstract}

\maketitle

\begin{quotation}
The finding of self-similar and self-affine traits, namely with multiple scaling exponents,
is presently as frequent as the verification of scale-dependence was in the early 20th century. The
identification of scale-invariant properties has been crucial to the accurate interpretation
and modeling of a long list of problems whereon climate and its dynamics rank high.
Despite being frequently studied, there is still much debate on the validity of
this concept in order to characterize the behavior of a time series as well as the importance of
each ingredient yielding scale invariance, namely dynamical dependencies and non-Gaussianity.
Moreover, the nonstationarity of most of the signals and the time-dependence of
each contribution to the scale-invariance properties have been systematically overlooked, particularly
in the case of the temperature anomaly in climatic time series.
\end{quotation}

\section{Introduction}

The interest of the scientific community in systems exhibiting self-similarity,
\textit{i.e.}, the property whereby a system is (approximately) equal to a part of itself,
has increased ever since Mandelbrot introduced the concept of fractality~\cite{mandelbrot}. In fact, aiming at analyzing the existence of
scale-invariant behavior, fractals have been applied to such diverse fields as
physiology and economics. Generically, scale invariance (or self-similarity) can be characterized
by a sole fractal (or Haussdorf) dimension or a spectrum of locally dependent exponents, the so-called multifractal
spectrum~\cite{mfbooks1,mfbooks2,mfbooks3}. Considering an abstract observable, $\mathcal{O}$,
scale invariance (self-similarity) features can be mathematically written as,
\begin{equation}
f\left( \left\{ \lambda \,\mathcal{O}^{z }\right\} \right) =\lambda ^{\alpha
\left( z \right) }\,f\left( \left\{ \mathcal{O}\right\} \right),
\label{selfsim}
\end{equation}
so that when $\alpha(z) $ is constant for all $z$, the system has a single scale invariant behavior.
In the case of a time series, the multifractal concept is prominently self-affine, \textit{i.e}, it
has different scaling properties in the ordinate ($\mathcal{O}$) and abscissa (time) directions.

In the fields of climate science and geophysics, fractality and scale invariance are popular
concepts as well. Moreover, the (multi-)fractal nature of meteorological times series and its connection to the 
widely discussed issue of climatic change has been
explored in several studies including instrumental \cite{kurnaz,alvarez} and paleoclimatic\cite{mf_dfa_appl3,bodri,karimova} temperature records and precipitation data\cite{valencia,rubalcaba}.
However, the large majority of the surveys on self-similarity in climate data have only been devoted to the analysis
of memory effects~\cite{dyer,bain,lovejoy,matyas,pelletier} in the form of
linear dependencies, \textit{i.e}, dependencies detected when the Pearson correlation function is computed.
In that case, the existence of memory in the data is quantified by the power spectrum exponent $\beta$, which is related to
a single value of $\alpha$ and that coincides with the Hurst exponent $H$. The value
of $\alpha$ relates to the power spectrum exponent $\beta$ via the Wiener-Khinchin relation and the fractal
dimension, $D$ reading $D = 2-H$. When $\alpha (z)$ is not constant, it is possible to go
beyond standard memory effects and explore a wider range of statistical and dynamical
properties.

In this manuscript we shed light on the multi-self-similar nature of a
well-known long meteorological time series: the anomaly of the
Central England Temperature (CET), which is the longest instrumental daily data set
available on this subject. The temperature anomaly is the difference between an element of the
time series and its long-term average, \textit{i.e.} the deviation of the daily temperature
from its annual cycle. Specifically, our work seeks to provide quantitative answers to the following questions: \emph{i)} To what
extent does the CET anomaly series show real multifractality? \emph{ii)} What is
the weight of each contributing mechanism to the multifractality we measure?
\emph{ii)} Does the relative weight of these contributions change in time?
If so, how and why do these changes occur? How much has been the relative magnitude of the changes?
Does this dynamics have to do with climate change?

The present manuscript is organized as follows:
in Sec. \emph{II} we provide a thorough description of the data analyzed and the
methods we employed; in Sec. \emph{III} we present our results concerning the
proposed questions, and in Sec. \emph{IV} we discuss
the results and present some conclusions we are able to draw from our analysis.

\section{Data and Methods}

\subsection{The Central England Temperature time series}
The Central England Temperature (CET) is the longest instrumental time series, starting in 1772.
Owing to its large span, it can be used to perform comparisons
between the pre-industrial era, during which the emission of greenhouse gases was low, and
contemporary times. That feature hints at nonstationary (trending) behaviour of the CET series, a trait that has been addressed in several prior works\cite{harvey,hannachi,kwon,tzhang,xzhang,franzke}.

\begin{figure}[tbh]
\begin{center}
\includegraphics[width=0.72\columnwidth,angle=0]{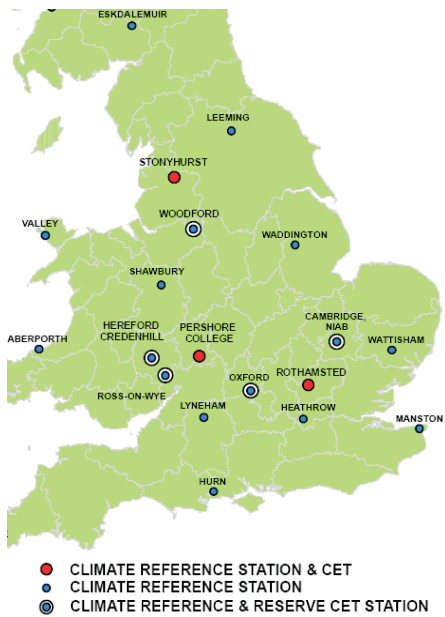}
\end{center}
\caption{Excerpt of Great Britain's map with climate stations locations
according to the legend shown in the figure. $\copyright $ Crown 2009, the
Met Office (Reproduced with permission.)}
\label{mapauk}
\end{figure}

The CET corresponds to a daily average of measured
temperatures at different climate stations close to the triangle defined by
the English towns of Bristol, Lancashire and London~\cite{manley1953,manley1974,parker1992,parker2005} (see Fig.~\ref{mapauk}).

The first compilation of a monthly series was presented by
Manley~\cite{manley1953,manley1974} covering the years from 1659 to 1973.
Afterwards, these data were updated to 1991 when Parker \emph{et al.} (1992)
calculated the daily series. Taking urban warming into account, the same data have been
adjusted by 0.1-0.3 Celsius degrees since 1974. From January 1878 until
November 2004, the data were recorded by the Met Office using the stations of
Rothamsted, Peershore College and Ringway and from the latter date onwards the
station of Stonyhurst replaced Ringway. At the same time, revised urban warming
and bias adjustments have been introduced in the data registered at Stonyhurst station.

The construction of the CET time series is a five-step procedure. In the two
first steps, the daily maximum and minimum temperatures (or the temperature
at specific hours) are averaged for specific station and the values from the stations
are combined. In the remaining stages, adjustments related to monthly
average, variance and urban warming are performed, resulting in a homogeneous
times series, as confirmed by previous studies \cite{parker2005,jones}. Further details about each step are described in Parker and Horton (2005)\cite{parker2005}, Section 3. The meticulous work in constructing and validating the CET series ensure both the reliability the accuracy of the data\cite{jones87,zorita}.

\begin{figure}[tbh]
\begin{center}
\includegraphics[width=0.95\columnwidth,angle=0]{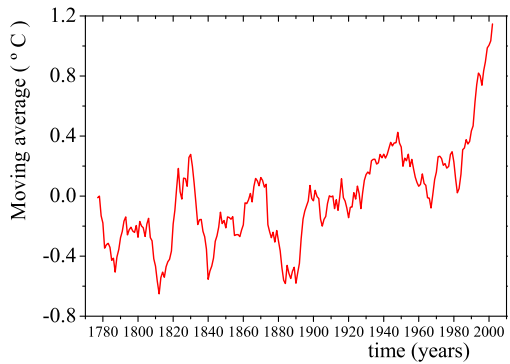}
\end{center}
\caption{Daily temperature anomaly averaged over a $11$-year sliding window (step size of one year) for the Central England Temperature (CET) time series.}
\label{movingaverage}
\end{figure}

In the present work, we analyze the daily temperature anomalies, which are the differences between the daily temperatures and the climatological annual cycle of temperature, which is defined, for each day, as the sum of the temperatures of this day carried out over all the years available, divided by its number. The data used in this work have been obtained from KNMI (The Royal Netherlands Meteorological Institute) Climate Explorer website
{\tt http://climexp.knmi.nl/}. The annual moving average of the CET anomaly time series is presented in Fig.~\ref{movingaverage}.

In performing this study, we follow a twofold approach: first, we analyze the
overall self-similar characteristics of the CET anomaly and then we
focus on the nonstationarities of the time series. In both cases we employ the same quantitative tools.
Specifically, in the latter approach we scan the daily time series, from $1772$ to $2007$, using a
$11$-year sliding window. Each window starts on the $1st$
January of year $i$ and ends on the $31st$ December of the year $i + 10$, resulting in window of $11\times 365$ points.
If $N$ is the number of years in the series, there will be $N-10$ ($=226$) windows, the centers of which are located at
the year $i+5$, resulting in a step size of one year. We have chosen to use $11$-years sliding windows because:
\emph{a)} it contains more than $4000$ data points in each window, which is enough for a sufficiently accurate estimation of a multifractal spectrum~\cite{serrano};
\emph{b)} it produces a sufficient number of windows for an appropriate assessment of the evolution of the scaling
properties of the system (the larger the windows, the smaller their number);
\emph{c)} it does not affect our spectra and thus our results since no time scale multiple of $11-$years was found
in the spectrum analysis;
\emph{d)} in using overlapping instead of nonoverlapping windows we are able to detect significant changes that otherwise would have passed unnoticed.

\subsection{Methods}

\subsubsection{Multifractal detrended fluctuation analysis - MF-DFA}

Methods of scaling analysis, such as the Detrended Fluctuation Analysis (DFA) have already been used in other climate studies~\cite{kurnaz}. Particularly, this method has been employed in the study of the evolution of the Hurst exponent of temperature time series~\cite{alvarez}. The DFA quantifies the relation between the variance of the accumulated sum of the signal and the size of the interval considered in the sum. Before computing the variance as a function of this time interval a detrending procedure, which seeks to remove pre-existing
nonstationarities, is applied to the signal. The DFA equally weighs large and small deviations and is thus unable to provide any information about the influence of their magnitudes. The method can be modified to check whether there exists an unequal contribution of large and small variations in the scaling behavior of a time series. This generalization is called multifractal detrended fluctuation analysis (MF-DFA)~\cite{mf-dfa}.

The MF-DFA is one of the most applied methods to determine the multi-scaling features of a time series~\cite{mf-dfa-appl1,mf-dfa-appl2,mf_dfa_appl3,mf-dfa-appl4,canberra,zhou-chem,zhou-fin}. We have opted to apply the MF-DFA rather than employ the Wavelet Transform Modulus Maxima (WTMM)~\cite{wavelet}
because the WTMM is prone to introduce artificial multifractality in a higher degree~\cite{polacos}. Additionally, MF-DFA gives better results for short signals~\cite{polacos}.

Considering a time series $\left\{ x\left( i\right) \right\} $ ($x$
represents the temperature anomaly in our case) composed of $N$ elements,
the MF-DFA consists of the following steps:

\begin{itemize}
\item Determine the profile $Y\left( i\right) $ of the cumulative deviation of the elements from the mean

\begin{equation}
Y\left( i\right) =\sum_{l=1}^{i}\left[ x\left(l\right) - \left\langle x\right\rangle \right],
\label{sum1}
\end{equation}
where $\left\langle \ldots \right\rangle$ represents the average over
elements and $i=1,...,N$.

\item Divide the new profile $Y\left( i\right) $ into $
N_{s}\equiv \mathrm{int}\left( \frac{N}{s}\right) $ nonoverlapping intervals of
equal size $s$.

\item Compute the trend of each  interval by a least-squares adjustment method, and
from that the variance for each segment $\nu =1,\ldots ,N_{s}$.

\begin{equation}
F^{2}\left( \nu ,s\right) =\frac{1}{s}\sum_{l=1}^{s}\left\{ Y%
\left[ \left( \nu -1\right) \,s+l\right] -y_{\nu }\left( l\right) \right\}
^{2},
\label{local}
\end{equation}%
where $y_{\nu }\left( l\right) $
represents a $p^{th}$-order fitting polynomial in the segment $\nu$.
The value of $p$ might be relevant to the results. For the CET anomaly time series we used polynomials of $2 ^{nd}$-order,
because using higher order polynomials does not significantly affect the multifractal spectrum.

\item Determine the fluctuation function of order $z$, $F_{z}\left( s\right) $ given by
\begin{equation}
F_{z}\left( s\right) \equiv \left\{ \frac{1}{N_{s}}\sum_{\nu =1}^{N_{s}}%
\left[ F^{2}\left( \nu ,s\right) \right] ^{z/2}\right\}
^{1/z},\qquad \forall _{z\neq 0},
\label{zscale}
\end{equation}
and
\begin{equation}
F_{z}\left( s\right) \equiv \exp \left\{ \frac{1}{2N_{s}}\sum_{\nu
=1}^{N_{s}}\ln \left[ F^{2}\left( \nu ,s\right) \right] \right\}
,\qquad z=0.
\label{zscale_0}
\end{equation}

Owing to the fact that the number of points $N$ is not generally a multiple of the scale $s$, the procedure described in the previous item may be also performed on the series with the order of the nonoverlapping segments reversed. In this case, the sum in the fluctuation function (Eqs. \ref{zscale} and \ref{zscale_0}) will have twice the number of terms and must be divided by two.

Equation ~(\ref{zscale}) presents the basic difference between the DFA and the MF-DFA methods. In the DFA method $z$ has only one value ($z=2$), while in the MF-DFA $z$ can have any real value. Negative values of $z$ decrease the influence of large values of $F^2 (s)$ on the fluctuation function, whereas positive values of $z$ decrease the influence of small values. The behavior of $F_z(s)$ for different values of $z$ reveals the impact of the different scales present in the data.

\item
Assess the scaling behavior of $F_{z}\left( s\right) $ in $\log -\log $ plots of $F_{z}\left( s\right) $ versus $s$
for each value of z. If the series $\left\{ x\left( i\right)\right\} $ shows multiscaling features then,
\begin{equation}
F_{z}\left( s\right) \sim s^{h\left( z\right) }.  \label{mf-dfa1}
\end{equation}
\end{itemize}

When the exponent $h(z)$ is negative or close to zero (which is not the case in our results), some of equations described in this section  must be slightly modified \cite{mf-dfa}.

If $z < 0$, small fluctuations generally lead to large values of $h (z)$, whereas large fluctuations are described by small values of $h (z)$. If $z > 0$, the opposite is true.

It is possible to connect the scaling exponent $h(z)$ in Eq. (\ref{mf-dfa1}), to the scaling exponent $\tau (z)$ of the the partition function in the standard multifractal formalism \cite{feder}. For multifractal measures this function scales with the size $s$ of the interval as
\begin{equation}
Z_{z}\left( s\right) \sim s^{\tau \left( z\right) }.  \label{mf-dfa2}
\end{equation}
Since that the fluctuation function $F_{z}\left( s\right) $ is related to the partition function $Z_{z}\left( s\right)$ by~\cite{mf-dfa}:
\begin{equation}
Z_{z}\left( s\right) \sim s^{-1}\left[F_{z}\left( s\right)\right]^z,  \label{mf-dfa3}
\end{equation}
then, according to Eqs.~(\ref{mf-dfa2})~and~(\ref{mf-dfa3}) we have,
\begin{equation}
\tau \left( z\right) =z\,h\left( z\right) -1.  \label{tau1}
\end{equation}%
The Legendre transform of this equation yields,
\begin{equation}
f\left( \alpha \right) =z\,\alpha -\tau \left( z\right),  \label{mf-dfa-tau}
\end{equation}
where is $\alpha$ is exponent mentioned in the Introduction (Eq.~\ref{selfsim}) and $f(\alpha)$ is
the dimension of the subset of the data that is characterized by $\alpha$.~\cite{mf-dfa}

Finally, we can relate the exponent $h \left( z\right) $ to the exponent $\alpha $ by
\begin{equation}
\alpha =h\left( z\right) +z\frac{dh\left( z\right) }{dz},
\label{mf-dfa-alfa}
\end{equation}
and
\begin{equation}
f\left( \alpha \right) =z\,\left[ \alpha -h\left( z \right) \right] +1.
\label{mf-dfa-falfa}
\end{equation}
For $z=2$, $h\left( 2\right) \equiv H$ corresponds to the Hurst exponent
customarily determined by methods such as the original $R/S$ ratio or the
original DFA~\cite{dfa,hurst} with $H= (\beta + 1)/2$, where $\beta $ is the power spectrum exponent
 we mentioned in the Introduction. Complementary, $z=0$ give us the support
dimension of the self-affine structure. In the case of a monofractal, $h\left( z\right) $ is independent of $z$.
This implies homogeneity in the scaling behavior and thus Eq.~(\ref{tau1})
reduces to $\tau \left( z\right) =z\,H -1$. Explicitly, there are only
different values of $h\left( z\right) $ if large and small fluctuations
scale differently. Lastly, we must stress we are not using the term fractal
in absolutely formal way. In fact, although in some cases $h(z)$ matches
the box-counting dimension, this exponent generally does not verify all the properties
of the Hausdorff-Besicovitch dimension of a signal.  For further details on this subject the reader is pointed to Ref. \onlinecite{feder}.

\subsubsection{Components of multifractality}

\label{methodscomponents}

There are two ingredients that introduce multifractality in a time series:
deviations from a scale dependent distribution and dependence between
the measurements that compose the time series. These factors
are traditionally assumed as independent and thus the total multifractality usually stems
from the superposition of both. This means we can attribute a certain value
$h_{\mathrm{NG}}(z)$ to the non-Gaussianity and the difference
$h(z)- h_{\mathrm{NG}}(z)$ to the dependence between measurements~\cite{note1} However, we must remember
that this is not entirely true. For instance, processes with
time-dependent standard deviation such as  ARCH-like~\footnote{ The acronym ARCH stands for AutoRegressive Conditional Heteroskedasticity } proposals (the so-called heteroscedastic models), the two elements of
multifractality are strongly dependent because the non-Gaussianity of the
time series originates from the existence of dependencies between elements that are linearly dependent in
the local variance (or squared volatility in financial jargon), tough the variable is itself uncorrelated~\cite{archmodels}. When the dependence in the
volatility is set to zero, the outcome is
a Gaussian series and
the two contributions to multifractality are one and the same. From Eqs.~(\ref{tau1})-(\ref{mf-dfa-falfa}),
we can see that a straightforward way of assessing the existence of multifractal features is
to compute the difference, $\Delta \alpha $, between the minimum and the maximum
values of the spectrum $f( \alpha )$. If the series is a mono-fractal we have $ \Delta \alpha = 0$, or else we should be able to separate the
contributions due to non-Gaussianity, $\Delta \alpha _{\mathrm{NG}}$, and to
the linear, $\Delta \alpha _{\mathrm{LD}}$, and the nonlinear dependencies,
$\Delta \alpha _{\mathrm{NL}}$, so that,
\begin{equation}
\Delta \alpha = \Delta \alpha _{\mathrm{NG}} + \Delta \alpha _{\mathrm{LD}}
+ \Delta \alpha _{\mathrm{NL}}.  \label{deltaalfa}
\end{equation}

\begin{figure}[tbh]
\begin{center}
\includegraphics[width=0.95\columnwidth,angle=0]{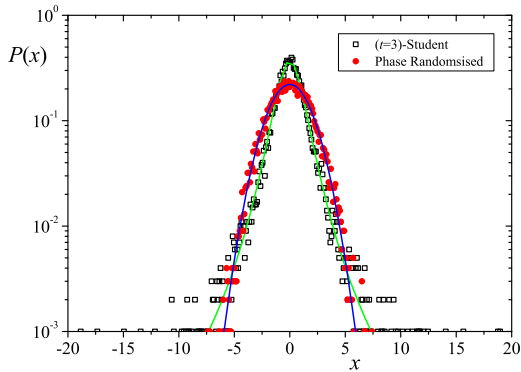}
\end{center}
\caption{(Color online) The $(\square )$ symbols were obtained from a
series of $10^4$ elements following a ($t=3)$-Student distribution (or a $%
(q=3/2)$-Gaussian with $\protect\sigma ^2 =3$) which was shuffled in order
to remove any possible dependence between the variables due to the
pseudo-random generator~\protect\cite{gentle} and the $(\bullet )$ symbols
represent the Gaussian of the surrogate time series
that is obtained by the phase randomization procedure.}
\label{ranproc}
\end{figure}

It is often assumed that linearities do not contribute to the multifractal width because they are traditionally
related to a single scale of dependence. However, previous results
on financial data~\cite{jef-csf,zhou-fin} have shown that these linear/weak dependencies do modify the MF-DFA spectrum.
Therefore, we shall not omit its contribution a priori.
Let us consider that the time series $\left\{ x\left( i \right)\right\} $ has
a non-Gaussian distribution and dependence between elements.
In performing an appropriate shuffling process, we are defining a new series with the same probability density function
as the original signal, but for which there is no dependence between the elements, because the memory was totally destroyed.
As proven in Ref.~\onlinecite{mf-dfa}, when we analyze the multifractal spectrum of the
shuffled surrogate, it will differ from the mono-fractal behavior, $\Delta
\alpha = \Delta \alpha_{\mathrm{(shuf)}} \neq 0$, only due to the non-Gaussianity of the probability
density function and thus,
\begin{equation}
\Delta \alpha _{\mathrm{NG}} = \Delta \alpha ^{\mathrm{eff}} _{\mathrm{(shuf)%
}},  \label{deltaNG}
\end{equation}
where  $\Delta \alpha ^{\mathrm{eff}}$ is the effective  $\Delta \alpha$ that
will be defined in the next paragraph. On the other hand, we can apply a procedure of
phase randomization in which we Fourier transform the series in the frequency
domain $\omega $ and replace the phases with uniformly distributed random values up to the first half of the transformed series
(excluding the phases of $\omega = -\pi,0$) and complete the surrogate series we use their conjugate values in the remaining of the series. Afterwards,
we Inverse Fourier transform the surrogate. The final result is a time series in
which the previously existing deviations from the Gaussian has been
destroyed (see an example in Fig.~\ref{ranproc}). Nonetheless, we can verify that
the power spectrum of the surrogate is the same as the original series, because we simply changed the phases of the Fourier Transform
of the time series whilst preserving its absolutes values.
Since the power spectrum describes the linear dependencies (or correlations), when we
evaluate the multifractality of the new series, $\Delta \alpha _{%
\mathrm{(rand)}} $, the only contribution originates in the linearities of
the system,
\begin{equation}
\Delta \alpha _{\mathrm{LD}} = \Delta \alpha _{\mathrm{(rand)}} ^{\mathrm{eff}}.
\label{deltaD}
\end{equation}

When we analyze the multifractal nature of a series that was both shuffled and phase
randomized (or vice-versa) we are expected to get $\Delta \alpha _{\mathrm{(shuf\& rand)}} = 0$.
However, in practice, $\Delta \alpha _{\mathrm{(shuf\& rand)}} \neq 0$.
These values different from zero (considering the error bars) are important
as they can be used to check the effect of the error introduced by the finiteness
of the time series and the method. This allows us to define the effective
$\Delta \alpha ^{\mathrm{eff}} $ as,
\begin{equation}
\Delta \alpha ^{\mathrm{eff}} _{(\ldots )} = \Delta \alpha _{(\dots )} -
\Delta \alpha _{\mathrm{(shuf\& rand)}},  \label{deltaeff}
\end{equation}
for the original, shuffled and random phase time series.

In addition, we assess the joint contributions of the non-Gaussianity and
the nonlinearities proceeding in the following way: instead of randomizing the phases of the
Fourier Transform ${\tilde{x}(\omega ) }$, we preserve them and assign a constant value to the
absolute value of $\tilde{x}(\omega ) $ keeping in mind that $| \tilde{x}(\omega )
|= |\tilde{x}(- \omega ) |.$ Acting in this way, we define a new surrogate of the noise
with a constant power spectrum, which is typical of a white noise, and with the remaining
dynamical features equal to the original signal because of the preservation of the phase. After
Inverse Fourier transforming, we compute the multifractal width $\Delta \alpha _{\mathrm{PP}}$,
the effective value of which is to be compared with the sum of $\Delta \alpha _{%
\mathrm{NG}}$ and $\Delta \alpha _{\mathrm{NL}}$.

In spite of the fact that the effect of time dependencies in
the multifractality is already taken into consideration in the
MF-DFA procedure (D stands for \emph{detrended}), it is worth mentioning that in order to apply the Wiener-Khinchin relation
the series must be stationary or at least close to it. With the
purpose to overcome the problem of the nonstationarity of the time
series for a proper evaluation of $S(\omega)$, which is only necessary to generate the surrogate series, we have applied a
high-pass filter ~\cite{highpass} to remover eventual nonstationarities (frequencies lower than $1/N$). Moreover, we have also used the Burg algorithm ~\cite{burg1,burg2}
for estimating $S(\omega)$ since it is very difficult to distinguish noise from
information in the standard FFT spectrum.
Nevertheless, we can still ask the following question: what are the actual advantages of a multifractal study?
Simple systems are traditionally characterized by the existence of a typical scale that implies an exponential dependence of quantities such as distribution, correlation and relaxation. For more complex systems, which are governed by nonlinear mechanisms, the existence of a typical scale is replaced with scale invariance relations that are depicted by power laws as that of Eq. (1). Therefore, when the system is ``weakly'' complex the data are described by a single exponent and the multifractal width vanishes. On the other hand, when the system is complex altogether, we have a series of power-law exponents defining the system and thus the multifractal width is different from zero. In other words, the larger $\Delta \alpha $, the more complex the system.

\section{Results}

\subsection{Overall results}

\begin{figure}[tbh]
\begin{center}
\includegraphics[width=0.95\columnwidth,angle=0]{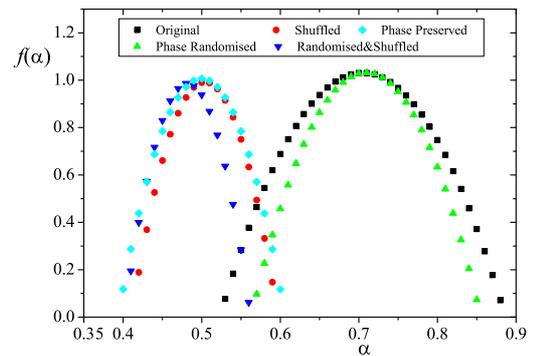}
\end{center}
\caption{(Color online) Multifractal spectrum $f(\protect\alpha )$ vs $%
\protect\alpha $ for the original CET anomaly and for the surrogates
generated by shuffling, phase randomizing, shuffling plus phase randomizing
and phase preservation.}
\label{cet-falfa}
\end{figure}

In this sub-section, we first analyze the time series as a single set.  The
scaling law  (Eq.~\ref{mf-dfa1}) is verified in the  range $s=27$ to approximately $s=16000$ days.
The results of the multifractal analysis of the CET anomaly time series and its
various surrogates described in the previous section are depicted in Fig.~%
\ref{cet-falfa}. For all the curves we verify the fat-fractal nature of the
series as the maximum of $f(\alpha )$ is equal to 1. In extrapolating the
values $\alpha _{\mathrm{min}}$ and $\alpha _{\mathrm{max}}$ at which the $f
(\alpha )$ curve intersects the $\alpha $ axis, we determined for the
original time series that $\Delta \alpha = 0.38$. For the shuffled plus
randomized surrogate we obtained $\Delta \alpha _{(\mathrm{shuf\& rand)}} = 0.16$
yielding an effective width $\Delta \alpha ^{\mathrm{eff}}= 0.22$.
Concerning the remaining surrogates we got the following
effective values $\Delta \alpha _{(\mathrm{shuf)}} ^{\mathrm{eff}} = \Delta
\alpha _{\mathrm{NG}} = 0.02$, $\Delta \alpha _{(\mathrm{rand)}}^{\mathrm{eff%
}} = \Delta \alpha _{\mathrm{LD}} = 0.14$ and $\Delta \alpha _{\mathrm{NL}}
= 0.06$. On the other hand, the computation of the multifractal width of the
phase-preserved surrogate yielded $\Delta \alpha _{(\mathrm{PP)}} = 0.06$. Taking into
account the error ($\pm 0.01$ in our analyses) the relation,
\begin{equation}
\Delta \alpha
_{\mathrm{PP}} \simeq \Delta \alpha _{\mathrm{NL}} +\Delta \alpha _{\mathrm{NG}},
\end{equation}
is verified.

The small contribution of the non-Gaussianity to the overall multifractal width
is easily understandable allowing for the quasi-Gaussian form of the CET anomaly
probability density function.

\begin{figure}[tbh]
\begin{center}
\includegraphics[width=0.95\columnwidth,angle=0]{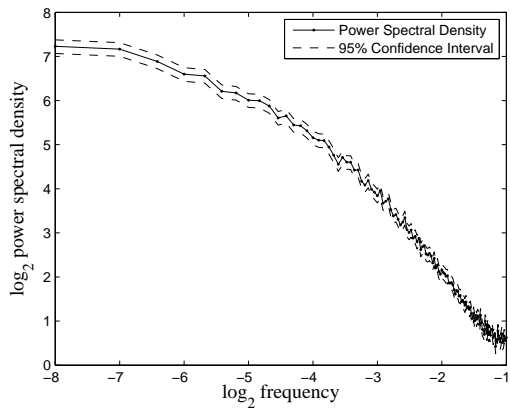}
\end{center}
\caption{(Color online) Power spectrum $S (\omega )$ vs $\omega $ for the CET
anomaly obtained using the Burg algorithm.~\cite{burg1,burg2} The dashed lines
represent the error margins ( 5 \% of the power spectrum value).}
\label{powerspectrum}
\end{figure}

\begin{figure}[tbh]
\begin{center}
\includegraphics[width=0.95\columnwidth,angle=0]{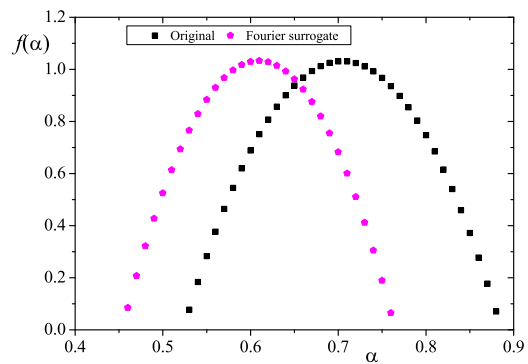}
\end{center}
\caption{(Color online) Multifractal spectrum $f(\protect\alpha )$ vs $%
\protect\alpha $ for the original CET anomaly and for the surrogate in which the power spectrum is kept.
In spite of being shifted one another, it is visible that the width of both spectra is are very similar.
Moreover, the multifractal width $\Delta \alpha _{(\mathrm{Fourier)}}$ lies within $\Delta \alpha _{(\mathrm{rand)}}$.}
\label{fig-fourier}
\end{figure}

These results, particularly the influence of the linear correlations
can be double-checked. Explicitly, we can define a
new surrogate which retains the power spectrum $S\left( \omega \right) $ (shown in Fig.~\ref{powerspectrum}) of
the CET anomaly but whose elements have a Normal distribution
and only exhibit linear dependencies between them. This is achieved by
first generating a series composed of a sequence independent and identically normally distributed
random variables that is, then, Fourier transformed. The amplitude of each element in the set $%
\left\{ \tilde{x}\left( \omega \right) \right\} $ is modified by multiplying
it by $\sqrt{S\left( \omega \right) }$. Thereafter, we apply the inverse Fourier
transform. The results of this procedure are depicted in
Fig.~\ref{fig-fourier}, which shows an effective width $\Delta \alpha
_{(\mathrm{fourier)}}^{\mathrm{eff}}=0.15$, completely compatible
with the result $\Delta \alpha _{\mathrm{LD}}$ presented before.

\subsection{Time dependence of the components of multifractality}

\begin{figure}[tbh]
\begin{center}
\includegraphics[width=0.95\columnwidth,angle=0]{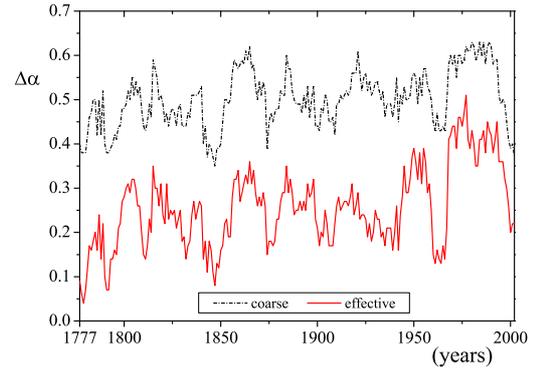}
\end{center}
\caption{(Color online) Multifractal width $\Delta \alpha$ of the Central England temperature
anomaly between 1777 and 2002,  coarse and effective. The $\Delta \alpha^{eff}$ is obtained
after removing finite size effects and systematic algorithmic error, by subtracting from the original
$\Delta \alpha$, the value of $\Delta \alpha_{(shuf\& rand)}$ as described in Eq.~(\ref{deltaeff}). Each curve follows the legend in the figure.}
\label{figuradosmfdfa}
\end{figure}

\begin{figure}[tbh]
\begin{center}
\includegraphics[width=0.95\columnwidth,angle=0]{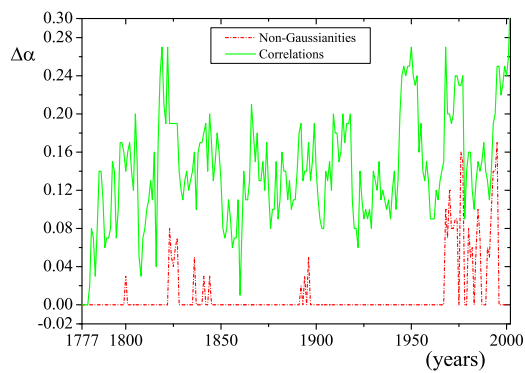}
\end{center}
\caption{(Color online) non-Gaussian and linear dependence (correlations) contributions to the
effective multifractal width $\Delta\alpha^{\mathrm{eff}}$ of the Central England temperature anomaly between
$1777$ and $2002$. Each curve follows the legend in the figure. The non-Gaussian character is
related to the width of the multifractal spectrum of the shuffled time series, while the
linear dependence is related to the width of the multifractal spectrum of the randomized time series.
All the calculations have been performed using the effective $\Delta \alpha$ given by Eq.~(\ref{deltaeff}).}
\label{figurasdoscomponentes}
\end{figure}

\begin{figure}[tbh]
\begin{center}
\includegraphics[width=0.95\columnwidth,angle=0]{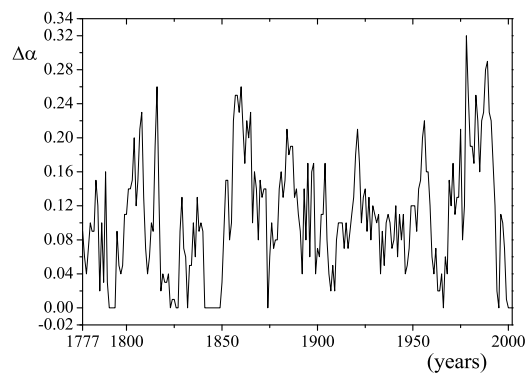}
\end{center}
\caption{(Color online) Nonlinear dependence contribution to the effective multifractal width
$\Delta\alpha^{\mathrm{eff}}$ of the Central England temperature anomaly between 1777 and 2002. This
contribution is measured by the width of the $f(\alpha)$ curve of the
surrogate time series with the same Fourier spectrum of the original one, but without
linear dependence. All the calculations have been performed using the effective $\Delta \alpha$
given by Eq.~(\ref{deltaeff}).}
\label{figuradocoupled}
\end{figure}

We now present the results obtained with 11-year sliding windows which allow
us to describe the dynamical features of multifractalitity in the CET anomaly.
The multifractal spectrum  has been fitted in range $16-1024$ days.
Fig.~\ref{figuradosmfdfa} shows how the multifractality has changed in time.
This multifractal dynamics can be fine-tuned by obtaining the effective multifractal
width $\Delta \alpha ^{\mathrm{eff}}$. The results of this refining are
striking. On average, the effective multifractality
represents just about $50 \% $ of the width obtained from
the original CET anomaly time series in Fig. \ref{figuradosmfdfa}. Concerning the elements of
multifractality, we can use Eq.~(\ref{deltaNG}) and Eq.~(\ref{deltaD}). Analyzing
the contribution of non-Gaussianity (Fig.~\ref{figurasdoscomponentes}), $\Delta \alpha _{NG}$, we see
that only $33$ out of the $226$ values of $\Delta \alpha ^{\mathrm{eff}} _{(shuf)}$ are
actually different from zero and consequently from a single-structure nature.
This implies an average contribution of merely 3\%, which once again is in accordance with
the quasi-Gaussian nature of the time series. The contribution linear dependencies, $\Delta \alpha _{LD}$,
which is obtained when we convert our series into a Gaussian series by means of the phase randomization procedure,
is of 59\% (Fig.~\ref{figurasdoscomponentes}). Adding
the two contributions we do not get 100 \%. This implies that the multifractality
introduced by nonlinearities corresponds to an average of 38\% (Fig.~\ref{figuradocoupled}).

\subsection{Detecting changes in climate}

\begin{figure}[tbh]
\begin{center}
\includegraphics[width=0.95\columnwidth,angle=0]{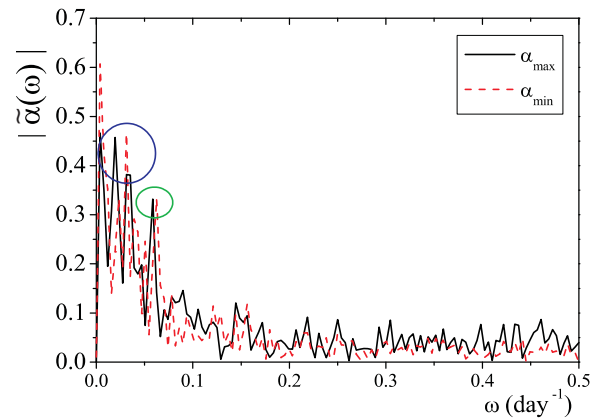}
\end{center}
\caption{(Color online) Amplitude of Fourier transform ($\tilde{\alpha }(\omega)$) of the maximum and the minimum
values of $\alpha(t)$, against frequency. The circles identify the local maxima described in the text.}
\label{alfasfourier}
\end{figure}

The results we have presented in the previous sub-section can be compared with the
results published by Berkes \emph{et al.}~\cite{royal}.
Using on Functional Data Analysis (FDA), Berkes \emph{et al.} pointed out a set
of statistically significant climate changes in the years of 1780, 1815,
1926 and 2007. Using a different method that the same authors named MDA, the climate changes
occurred in 1780, 1808, 1850, 1926, 1992 and 2007.
Our results, show changes in the non-Gaussianity around 1800, followed by
the flurried periods 1820-1829, 1834-1845, 1891-1897 and a last and standing
disturbed period from 1966 on (Fig.~\ref{figurasdoscomponentes}). Although Berkes \emph{et al.}\cite{royal} analyzed climate variations by measuring
quantities different from the multifractal characteristics determined here,
changes in the dynamics of a given observable can be identified using
very different quantities the variations of which are related to the dynamics of this observable.
Therefore, it is legitimate to consider this concurrence indicative of the robustness of past changes in the leading dynamical mechanism of the temperature anomaly.
Although we could not observe any change in 1926 or thereabouts in Fig.~\ref{figurasdoscomponentes}, we noticed that
there is a sharp peak close to this date in Fig.~\ref{figuradocoupled}. It is
worth emphasizing that Berkes \emph{et al.}~\cite{royal} to their methods, especially the
latest one, as a ``mere modeling assumption that is useful in identifying patterns of
change in mean temperature curves''.
Furthermore, we perceive that the broader and larger nonlinear and non-Gaussian contributions to
overall multifractality start in the mid 1960s and persist until the late 1990s. A pattern
of successive positive values in the annual average of the CET
anomaly was reported by Parker \emph{et al.} (1992) in this
period~\cite{parker1992} and encloses a global change of phase that is
well-documented in the climate literature~\cite{ipcc}. Last but not least, we have
observed that the peaks verified in the $\Delta \alpha _{\mathrm{NG}}$ (Fig.~\ref{figurasdoscomponentes}) concur
with the emergence of kurtosis excess when statistical moments are
surveyed~\cite{aguasdelindoia}. Recently, results by Hansen~\emph{et~al.}\cite{hansen} in a global climate change study
using a standard statistical approach indicated
that the temperature anomaly in the last decades exhibits non-Gaussian distributions.
Motivated by the time dependence of the scaling properties of the CET anomaly, we analyzed the spectra $\alpha
_{\mathrm{min}}$ and $\alpha _{\mathrm{max}}$ (effective values)
(Fig.~\ref{alfasfourier}). The maxima in the amplitude of both
spectra agree with decadal oscillations. Namely, we found for
$\alpha _{\mathrm{min}}$ maxima at $16^{-1}$ and $32^{-1}$ year$^{-1}$
and for $\alpha_{\mathrm{max}}$ maxima at $17^{-1}$, $34^{-1}$ year$^{-1}$ with the former having an extra
peak at $55^{-1}$ year$^{-1}$.  These decadal oscillations are not present when the width $\Delta \alpha$ is analyzed.
One reason for that stems from the fact that both $\alpha _{\mathrm{max}}$ and $\alpha _{\mathrm{min}}$ show very close peaks
in the spectrum and thus when the difference between them is considered the peaks are not perceived, which is similar to
what occurs with two co-evolving quantities.

\section{Final remarks and outlook}
In this manuscript we studied the multifractal features of the CET anomaly.
This was carried out by considering the entire time series and by scanning the time series with
11-year sliding windows. Our results show that the main component of the
multifractal structure of this time series is due to the linear dependencies in
the dynamics, followed by nonlinear dependencies and with a residual
contribution of the deviations from the Gaussianity. Nonetheless, when we analyzed how the
multifractality of the CET anomaly changes in time, we verified that until 1950 the non-Gaussian contribution appears in dates that are very close to the dates associated with climate changes, as presented in a previous study using Functional Data Analysis~\cite{royal}. From the 1960s onwards the contribution of the non-Gaussianity
becomes significant for most of the time, a result that matches a recent finding of temperature anomalies significantly
beyond the $3 \, \sigma $ criterion that characterizes the invalidity of the Gaussian~\cite{hansen}.

Finally, although the scope of the present work concerns the quantitative description of the multifractal properties of the CET anomaly, our analysis can have direct implications in modeling as well.
Our results indicate that the temperature variability modeling must take into account
the multifractal nature of the temperature time series, and also the contribution of each ingredient for the multifractality.
A tentative dynamical scenario is that wherein we resort to the statistical mechanics
relation between temperature and standard deviation, \emph{i.e.}, the temperature is proportional
to the standard deviation, and consider a model inspired in cascade models of the latter quantity
for turbulent fluids such as those introduced in Refs.~\onlinecite{meneveau,castaing,arneodo}.
Instead of considering a cascade process in the standard deviation of the quantity under analysis,
we reinterpret those models considering a cascade process for the average value of the observable,
which in this case is the temperature anomaly.
In this approach, the average value over a certain scale $\ell $
comes from the multiplicative process,
\begin{equation}
\mu _{\ell }\left( t\right) =\prod\limits_{i=0}^{n-1}f\left( i\rightarrow
i+1\right) \,\mu _{L},
\end{equation}
with $\mu _{L}$ representing the average over a reference period $L$, e.g. $%
L=11$ years, and $f$ representing the fraction of measure (average) passing
from a earlier generation (time scale) to the subsequent. In this case, the
greater $i$, the smaller $\ell $, i.e., $i=0$ corresponds to $\ell =L$ and $\ell =n$ yields the scale of the temperature anomaly we are intend  to describe. Consequently, the
temperature anomaly, $\xi $, would be equal to,
\begin{equation}
\xi _{\ell }\left( t\right) =\varepsilon \left( t\right) +\mu _{\ell }\left(
t\right) ,
\end{equation}
where $\varepsilon $ represents a Gaussian independent and identically
distributed noise with zero average and appropriate standard deviation. A skew distribution of $\mu _{L}$ will lead to a
skew distribution of $\xi $ as well. Concomitantly, in order to represent the
slight kurtosis we can consider a further contribution, $\zeta $, coming
from a multiplicative noise process,
\begin{equation}
\zeta \left( t\right) =\eta \left( t\right) \,\sigma \left( t\right)
\,h\left( t\right) ,
\end{equation}
with $\sigma \left( t\right) $ being a function of past values of $\xi $ and
$\zeta $ in a heteroscedastic way (see for details~\onlinecite{garch,andersen})
and $\eta $ another Gaussian noise not correlated with $\varepsilon $. The $\zeta $ contribution is modulated by a step function $h\left(
t\right) =\Theta \left[ t-\left( \Upsilon -\Delta \Upsilon \right) \right]
\Theta \left[ \Upsilon +\Delta \Upsilon -t\right] $ where the center of the
intervals come from a shot noise following a frequency related to Fig. (\ref{alfasfourier}),
\textit{i.e.}, a random process that has nonzero values at an average rate, $\lambda$,
which belong to the Poisson class of stochastic processes.
Mathematically, the probability of having a contribution from $\zeta $
arising from a perturbation centered at time $\Upsilon $ is given by,
\begin{equation}
p\left( \Upsilon \right) \varpropto \sum\nolimits_{i}\delta \left( \Upsilon
-\Upsilon _{i}\right) \,,
\end{equation}
and the total temperature anomaly will thus correspond to the sum of $\xi
_{\ell }\left( t\right) $ and $\zeta \left( t\right) $.

Alternative models can be presented, namely for mimicking the fluctuations of the
temperature anomaly instead of the temperature itself. It is simple to find one quantity
after the other bearing in mind their relation and statistical properties as in happens
in other problems such as fluid turbulence and price dynamics in financial markets.

One the other hand, still in the context of the empirical analysis, the present study can be
further expanded into the statistical and dynamical analysis of yearly $\Delta \alpha $ fluctuations,
especially its probability density function and the existence of a mixture dynamical regimes by the application
of nonparametric testing~\cite{dks}.

\subsubsection{Acknowledgements}
We would like to thank the Met Office for kindly providing us with the map in Fig.~\ref{mapauk}.
AMG has received funding from CNPq (Brazilian National Council for
Scientific and Technological Development), and from the EU Seventh Framework Programme (FP7/2007-2013) under Grant Agreement nr 212492
(CLARIS LPB. A Europe-South America Network for Climate Change Assessment and
Impact Studies in La Plata Basin). JdS acknowledges CNPq, PETROBRAS$/$FUNPAR-UFPR, Setor
de Ci\^{e}ncias Exatas - UFPR and Setor de Ci\^{e}ncias da Terra - UFPR for support and S. H. S. Schwab,
F. Mancini, F. J. F. Ferreira, are also thanked for valuable helping and encouragement. SMDQ thanks
the LABAP laboratory of Universidade Federal do Paran\'{a} for the warm hospitality during his visits
to the institution  sponsored by CNPq and PETROBRAS$/$FUNPAR-UFPR and the European Commission through the Marie Curie Actions FP7-PEOPLE-2009-IEF
(contract nr 250589).


\begin{thebibliography}{}
\bibitem{note1} Throughout the text we employ the term non-Gaussian in the broad sense, \emph{i.e.}, we use it to refer to a distribution
that does not have a defined scale as the Gaussian and which is a established terminology for the absence of scale.
\end{thebibliography}
\end{document}